\documentclass[prl,twocolumn,unsortedaddress,superscriptaddress,floatfix]{revtex4}

\usepackage{graphicx}
\usepackage{dcolumn}
\usepackage{bm}
\usepackage{amsmath}

\begin{document}

\title{Direct observation of oxygen superstructures in manganites}

\author{S. Grenier} \affiliation{Institut N\'eel, CNRS \& Universit\'e
Joseph Fourier, BP 166, F-38042 Grenoble Cedex 9, France}

\author{K. J. Thomas} \author{J. P. Hill} \affiliation{Condensed Matter
Physics and Materials Science Dept., Brookhaven National Laboratory,
Upton, New York 11973, USA}

\author{U. Staub} \author{Y. Bodenthin} \author{M. Garc\'ia-Fern\'andez}
\affiliation{Swiss Light Source, Paul Sherrer Institut, 5232 Villigen,
Switzerland}

\author{V. Scagnoli} \affiliation{European Synchrotron Radiation
Facility, BP 220, F-38043 Grenoble Cedex 9, France}

\author{V. Kiryukhin} \affiliation{Department of Physics and Astronomy,
Rutgers University, Piscataway, New Jersey 08854 USA}

\author{S-W. Cheong} \affiliation{Department of Physics and Astronomy,
Rutgers University, Piscataway, New Jersey 08854 USA}

\author{B. G. Kim} \altaffiliation{Department of Physics, Pusan National
University, Pusan 609-735, Korea}\affiliation{Department of Physics
and Astronomy, Rutgers University, Piscataway, New Jersey 08854 USA}
 
\author{J. M. Tonnerre} \affiliation{Institut N\'eel, CNRS \&
Universit\'e Joseph Fourier, BP 166, F-38042 Grenoble Cedex 9, France}

\date{\today}

\begin{abstract} 
We report the observation of superstructures associated with the oxygen
$2p$-states in two prototypical manganites using x-ray diffraction at
the oxygen $K$-edge. We determine the nature of the orderings and
discuss our picture with respect to novel theoretical models. In the
stripe order system Bi$_{0.31}$Ca$_{0.69}$MnO$_3$, hole-doped O states
are found to be orbitally ordered, at the same propagation vector as the
Mn orbital ordering, but no evidence is found to support a picture of
oxygen charge stripes at this periodicity. In
La$_\frac{7}{8}$Sr$_\frac{1}{8}$MnO$_3$, we observe a $2p$ charge
ordering described by alternating hole-poor and hole-rich MnO planes
that is consistent with recent predictions.
\end{abstract}

\maketitle

Fascinating macroscopic properties may emerge from the electronic and
magnetic orderings that occur in 3\emph{d} metal oxides doped with
charge carriers. In doped cuprates, charge and spin stripes could be
relevant to high-T$_C$ superconductivity \cite{tranquada:nature1995},
and the colossal magnetoresistance of doped manganites is fundamentally
related to the stability, in a magnetic field, of an ordering of
polarons together with an ordering of charge, orbital and magnetic
moments on the Mn atoms \cite{CMRManganites}. The relationship between
the orderings and the transport properties has been adressed both
theoretically and experimentally concentrating on the $3d$ metal atoms,
often neglecting the ligands' degrees of freedom. However, because the
active states are not ionic 3\emph{d} metal orbitals, but, rather
hybridized metal 3\emph{d} - oxygen 2\emph{p} states, an ordering of the
oxygens may be related to the ordering of the metals. In cuprates, the
role of the oxygen atom is now always considered and, taking a recent
example, the ordering of hole-doped oxygens was directly observed and
discussed in the context of the stripes \cite{rusydi:016403}.

In manganites, an electronic ordering on oxygens has not been detected
yet although recent experiments suggested a crucial role of the O atoms
in the polaronic, electronic, and magnetic orderings of the Mn atoms. A
relation between conductivity and the O electronic configuration was
identified \cite{abbate}; spectroscopic studies indicated a small charge
disproportionation on Mn atoms instead of an ionic charge ordering
(Mn$^{4+}$/Mn$^{3+}$) \cite{subias:PhysRevB.56.8183}, and a
crystallographic study proposed that the oxygen atom bridges an electron
between two Mn atoms of nearly the same valence, coupling them
magnetically \cite{daoud-aladinePRL02}.  At the heart of these results
is the explicit role of the $2p-3d$ hybridization in determining the
magnetic exchange couplings and conductivity, the traditional picture
being the one where the atomic $3d$ orbitals define exchange magnetic
pathways between Mn atoms, the oxygen $2p$ states being filled.

Meanwhile, new theoretical models beyond the Mn-centered description
also suggested that oxygen superstructures play an important role in
determining the ground states.  Ferrari \emph{et al.}
\cite{ferrari:227202} explained the insulating behavior of the
charge/orbitally orded half-doped system by an ordering of O holes -
``oxygen stripes'' of varying valences (incidentally, having the same
periodicity as the Mn orbital ordering). Efremov \emph{et al.}
\cite{efremov:naturemat} proposed that the charge ordering could
correspond to a charge density wave (CDW) that is either Mn-centered or
O-centered; they then show that an intermediate CDW (an intermediate
O-Mn charge transfer) would potentially give both ferromagnetism and
ferroelectricity. Finally, Volja \emph{et al.}  \cite{volja:0704.1834v1}
insisted on the relevance of the Mn charge ordering concept, if the
hybridization with the hole-doped oxygen atoms is properly considered.

Experiments that directly probe the oxygen contribution to the ordered
states provide essential feedback to these theoretical models. In this
Letter, we show that O $K$-edge resonant diffraction spectra reveal
oxygen superstructures in two manganites. We first report the direct
observation of the \emph{orbital} ordering of hole-doped, hybridized, O
$2p$-states in a Mn-centered, Mn-orbitally ordered manganite,
Bi$_{1-x}$Ca$_{x}$MnO$_3$ $(x=0.69)$. Second, we report the
\emph{charge} ordering on the oxygen in
La$_\frac{7}{8}$Sr$_\frac{1}{8}$MnO$_3$; our observations confirm the
stacking of alternate hole-rich and hole-poor planes. In both cases, the
O states are directly related to the Mn states, emphasizing the
hybridized nature of the active states involved in the electronic
orderings.

\begin{figure}
\includegraphics[width=0.95\columnwidth]{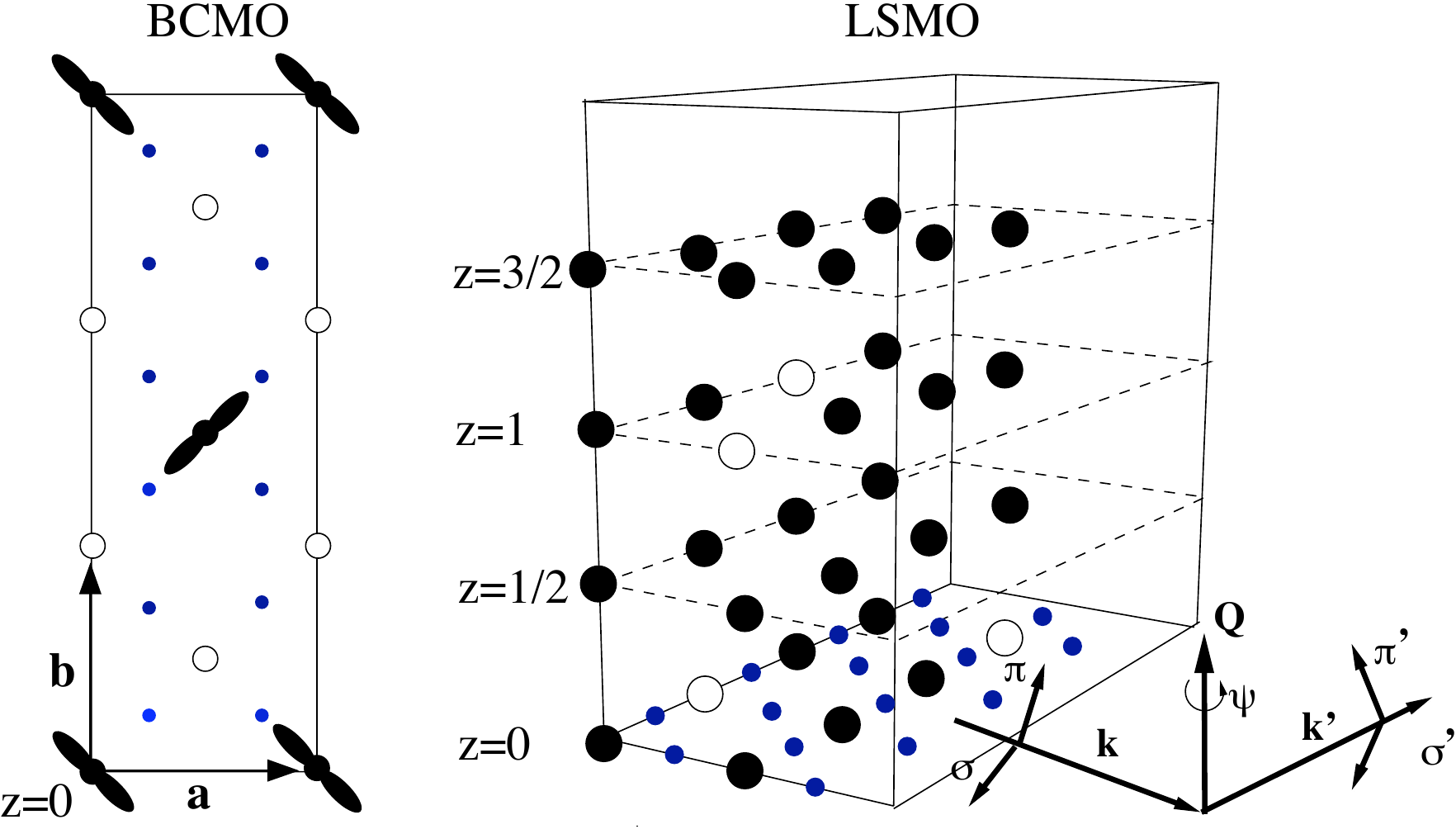} 
\caption{\label{fig1}
Simplified orderings in BCMO and LSMO. Closed circles represents
polaronic Mn$^{3+}$ atoms, open circle are Mn$^{4+}$ and small circles
are the O sites; the orbital ordering in BCMO is also shown. The
structure in the a-b plane of BCMO repeats in adjacent planes, while
LSMO planes alternate between being hole-rich (z=0 and 1) and hole-poor
(z=1/2 and 3/2) \cite{geck:236401}; for clarity the oxygens are
represented in the first plane only. The scattering geometry is
indicated, $\psi$ is the azimuthal angle, $\sigma$ and $\pi$ are the two
polarization components.}
\end{figure}

Bi$_{0.31}$Ca$_{0.69}$MnO$_3$ (BCMO) is representative of the
Mn-centered orbitally ordered CDW phase, which is usually described as
Mn stripe order.  This phase is robust in manganites, from half-doping
to $x=0.8$ \cite{CMRManganites}. The CDW exhibits polaronic sites
(Mn$^{3+}$) placed as far as possible from each other, with an antiferro
order of the valence orbital, while Mn$^{4+}$ sites are isotropic
\cite{grenier:085101} (Fig. \ref{fig1}). As the doping increases, the
polaron density diminishes, and a larger orbital ordering periodicity is
obtained ($\Lambda\approx 17$ \AA\ for BCMO), which is large enough to
diffract soft x-rays near the O $K$-edge ($\lambda=23.4$ \AA\ at 530
eV). BCMO orders below $T_{OO}=250$ K.
The second system, La$_\frac{7}{8}$Sr$_\frac{1}{8}$MnO$_3$ (LSMO),
exhibits several phase transitions upon cooling. At $T_{OO1}\approx 280$
K, a polaronic insulating orbital ordering takes place. Then, the system
becomes a metallic ferromagnet at T$_C\approx 180$ K and finally
stabilizes below $T_{OO2}\approx 150$ K as an insulating ferromagnet. It
was recently demonstrated that the OO1 to OO2 transition corresponds to
a rearrangement of the polarons in a larger unit cell
\cite{geck:236401}. Here, we directly probed the electronic ordering
that underlies the polaron rearrangement allowing the test of two recent
models for the electronic structures (described below).

The BCMO compound was prepared by the flux technique and LSMO by the
floating zone technique; both samples were characterized by hard x-ray
diffraction. The crystals were cut and polished with the normal to the
surface along $[010]$ for BCMO, and along $[001]$ for LSMO.  We refer to
the P$bnm$ orthorhombic notations for the unit cell.

The RXD data were collected at the SIM beamline of the Swiss Light
Source using the RESOXS chamber \cite{JaouenRESOXS}. The experiments
were performed as in conventional diffraction but with the incident
energy varied through the O $K$-edge whilst moving the diffractometer
angles to maintain the diffraction condition. The O $K$-edge corresponds
to the photoelectric transition of a 1\emph{s} electron to unoccupied
orbitals $\psi_n$. This transition is accounted for in the scattering
factor by the overlap integral
$\hat{\varepsilon}\cdot\int\psi^*_n(\textbf{r})\textbf{r}\psi^\text{O}_{1s}(\textbf{r})d\textbf{r}$,
$\hat{\varepsilon}$ being the photon polarization. The resonant and the
Thomson scattering factors are then weighted by the crystallographic
phase-factor at the selected Bragg reflection, $F(Q,\omega)=\sum_j
f_j(Q,\omega) e^{i\textbf{Q} \cdot \textbf{r}_j}$, $\omega$ being the
energy of the photon and the energy between $\psi_n$ and
$\psi^\text{O}_{1s}$ \cite{Materlik}. The resonant reflection provides
evidence of the periodicity of near-$\varepsilon_f$ states, and sweeping
the incident photon energy through the O $K$-edge, therefore effectively
probes the spectrum of unoccupied states $\psi_n$ from and above the
Fermi level ($\varepsilon_f$) at this periodicity.  The unoccupied
orbitals probed must not have an even symmetry relatively to the oxygen
$1s$ orbital $\psi^\text{O}_{1s}$ (which is even), as easily seen from
the overlap integral. Near $\varepsilon_f$, those may be O $2p$'s, or
orbitals from nearby atoms with some projection on the oxygens
(\emph{e.g.}  $3d$), or orbitals made of a hybridization of the two. We
also measured azimuthal scans (record of the reflection intensity while
rotating the crystal in a plane normal to \textbf{Q}) which allow one to
rotate the polarization of the photon relatively to the crystal thereby
probing different directions of the unoccupied states
($\psi_{2p_x}$,...). We could select the two incident polarizations,
namely $\sigma$ or $\pi$, perpendicular to and within the scattering
plane, respectively (Fig. \ref{fig1}), the detector integrating photons
of both polarizations.

\begin{figure}
\includegraphics[width=0.8\columnwidth]{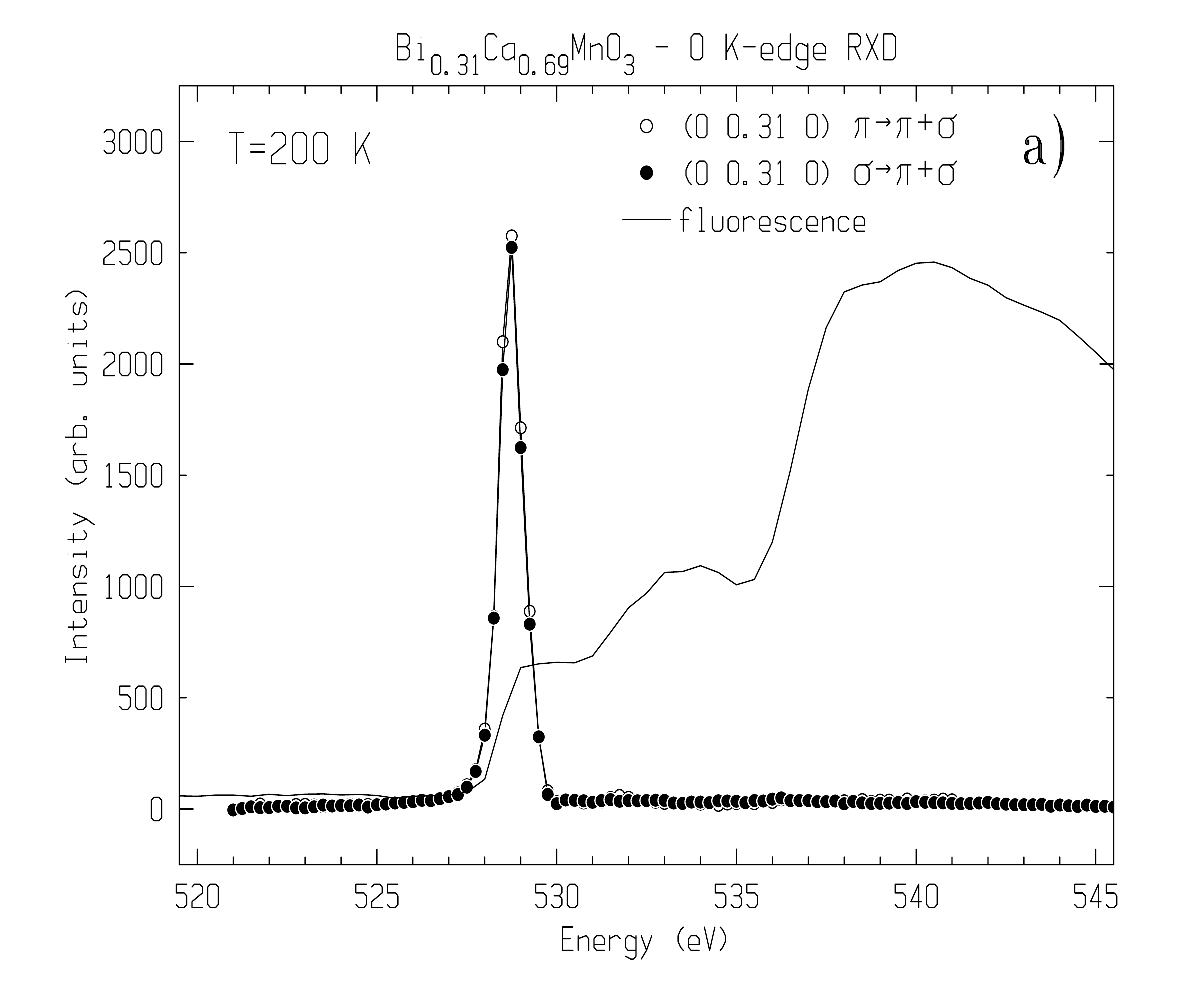}
\includegraphics[width=1\columnwidth]{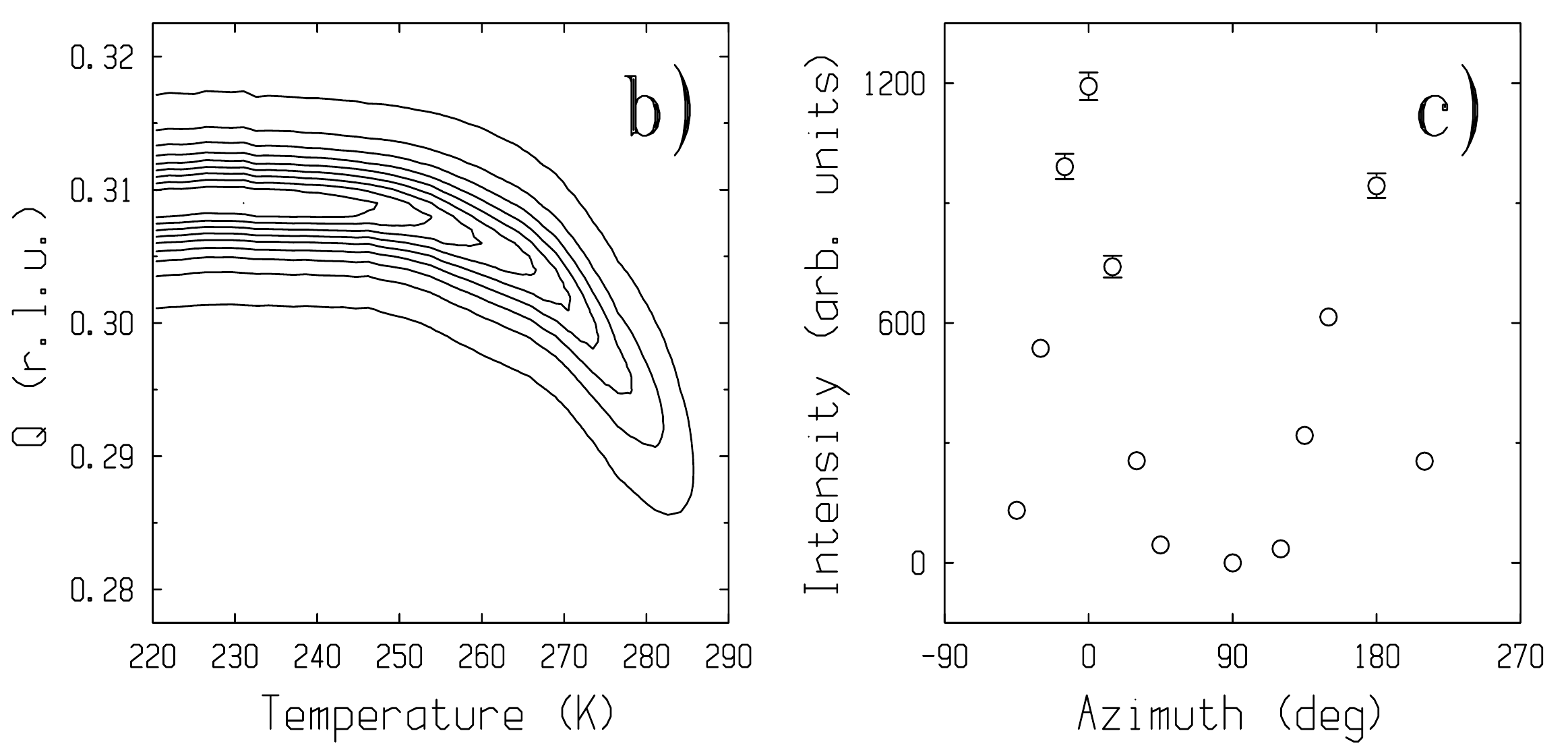}
\caption{\label{fig_BCMO} a) The $(0\ 0.31\ 0)$ superstructure
reflection, corrected from fluorescence but not for self-absorption;
nevertheless, the intensity is zero off the O $K$-edge. b) Temperature
dependence of the peak position at the phase transition. c) Azimuthal
dependence displaying a two-fold symmetry with a maximum with the
incident polarization along \textbf{a}, and zero along \textbf{c}, taken
at 528.75 eV, T=200K.}
\end{figure}

Fig. \ref{fig_BCMO}a shows the O $K$-edge RXD spectrum at the $(0\ 0.31\
0)$ superstructure reflection we found in BCMO, for $\pi$ and $\sigma$
incident polarizations. No scattering is observed outside a narrow
energy range (1.1 eV) at the edge of the absorption spectrum (given by
the fluorescence yield). We observed equal spectra for the two incident
polarizations. As has been discussed elsewhere \cite{grenier:085101}
this implies that the scattering process is one in which the
polarization of the photon is rotated, of the type
$\sigma\rightarrow\pi$ or $\pi\rightarrow\sigma$, that is, with zero
$\sigma\rightarrow\sigma$ and $\pi\rightarrow\pi$ scattering. The
incommensurability of this reflection is observed to vary with
temperature, stabilizing at about 0.31 r.l.u., below $T_{OO}$
(Fig. \ref{fig_BCMO}b). A final observation, from Fig. \ref{fig_BCMO}c,
is the extinction of the reflection when the polarization of the
incident photon is perpendicular to the ordered $(a,b)$ plane.

The first observation indicates a resonant superstructure reflection
from an ordering of the very first unoccupied states from
$\varepsilon_f$ projected on the O atoms ($\psi^O_{1s}$), with the
propagation $Q=0.31$ r.l.u, corresponding to the periodicity of the Mn
orbital ordering \cite{grenier:085101}. Following earlier analysis of
the absorption data \cite{Ju.PhysRevLett.79.3230}, the comparison with
the fluorescence yield allows us to identify the ordered states as the
$2p$ states hybridized with the $3d$. The narrow spectral width observed
reflects the width of this hybridized band. The second observation
demonstrates that there is no charge ordering on O atoms \emph{at this
wave vector}. Indeed, a CDW of this periodicity would scatter in the
$\sigma\rightarrow\sigma$ channels, which is absent on this reflection
to within the uncertainty of our measurements. Furthermore, a charge
order scatters at all azimuthal angles \cite{Grenier04,Garcia01}, at
odds with the observation of zeros of the signal. These three
observations prove that the O $2p$ states are hole-doped and show
orbital ordering, and not charge ordering at this propagation
vector. Therefore, the data exclude a direct extension to this high
doping of the model of oxygen stripes at the periodicity of the Mn
orbital ordering \cite{ferrari:227202}.

The observation of $2p$ orbital ordering makes intuitive sense in light
of the $2p-3d$ hybridization and the earlier observation of this same
orbital ordering, but projected on the Mn atoms ($\psi_{2p}$), at the Mn
$2p \leftrightarrow 3d$ $L$-edge \cite{grenier:085101}. Specifically,
these states are centered on the polaronic Mn atoms, with a
configuration depicted in Ref. \cite{yin:116405,volja:0704.1834v1}
clearly exhibiting the unoccupied $2p$ ordering on the four oxygens
surrounding the polaronic Mn. In agreement with this model, our results
clearly demonstrate the charge transfer nature of the compound and the
coupling between the O and Mn orbital ordering.

\begin{figure}
\includegraphics[width=0.8\columnwidth]{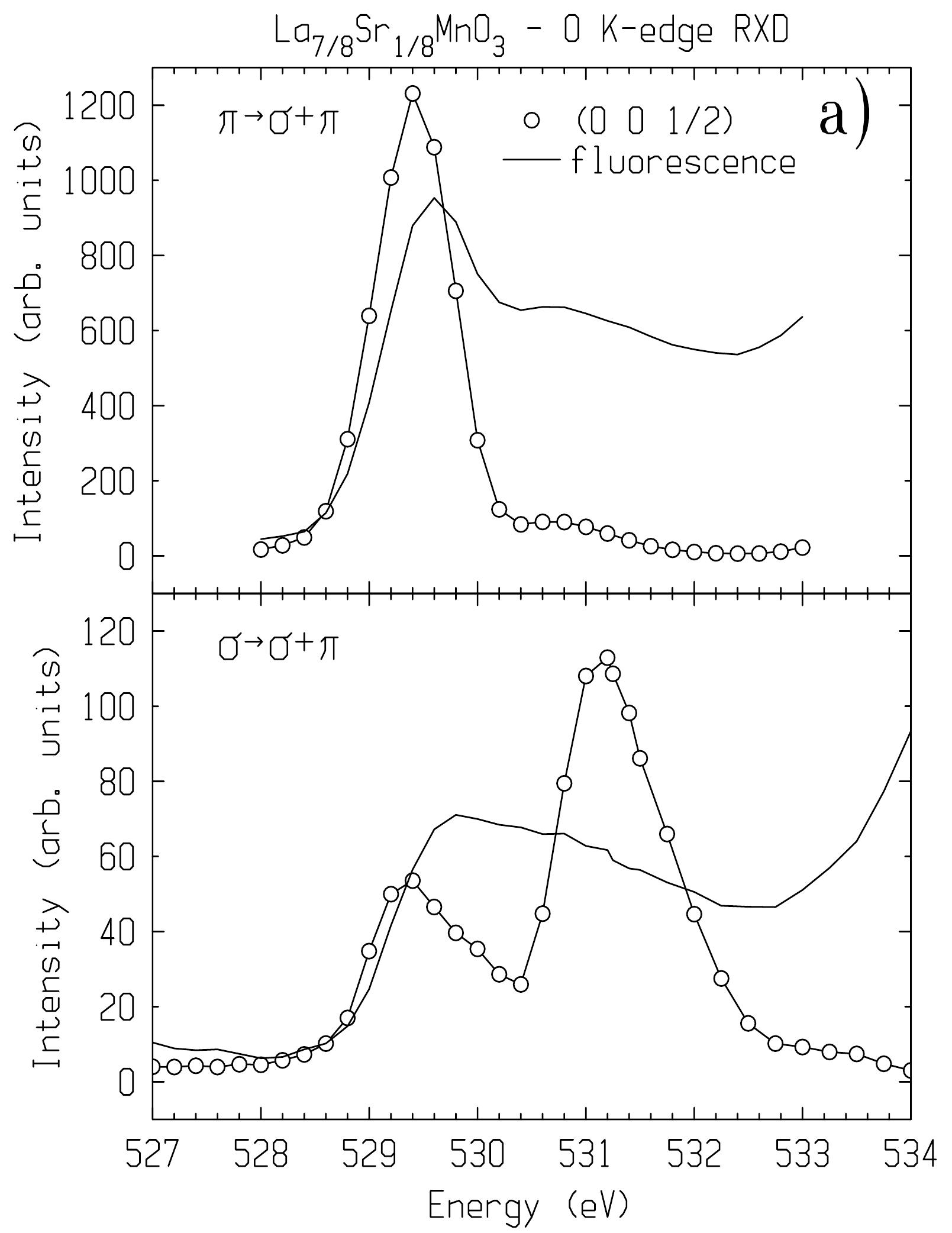}
\includegraphics[width=0.95\columnwidth]{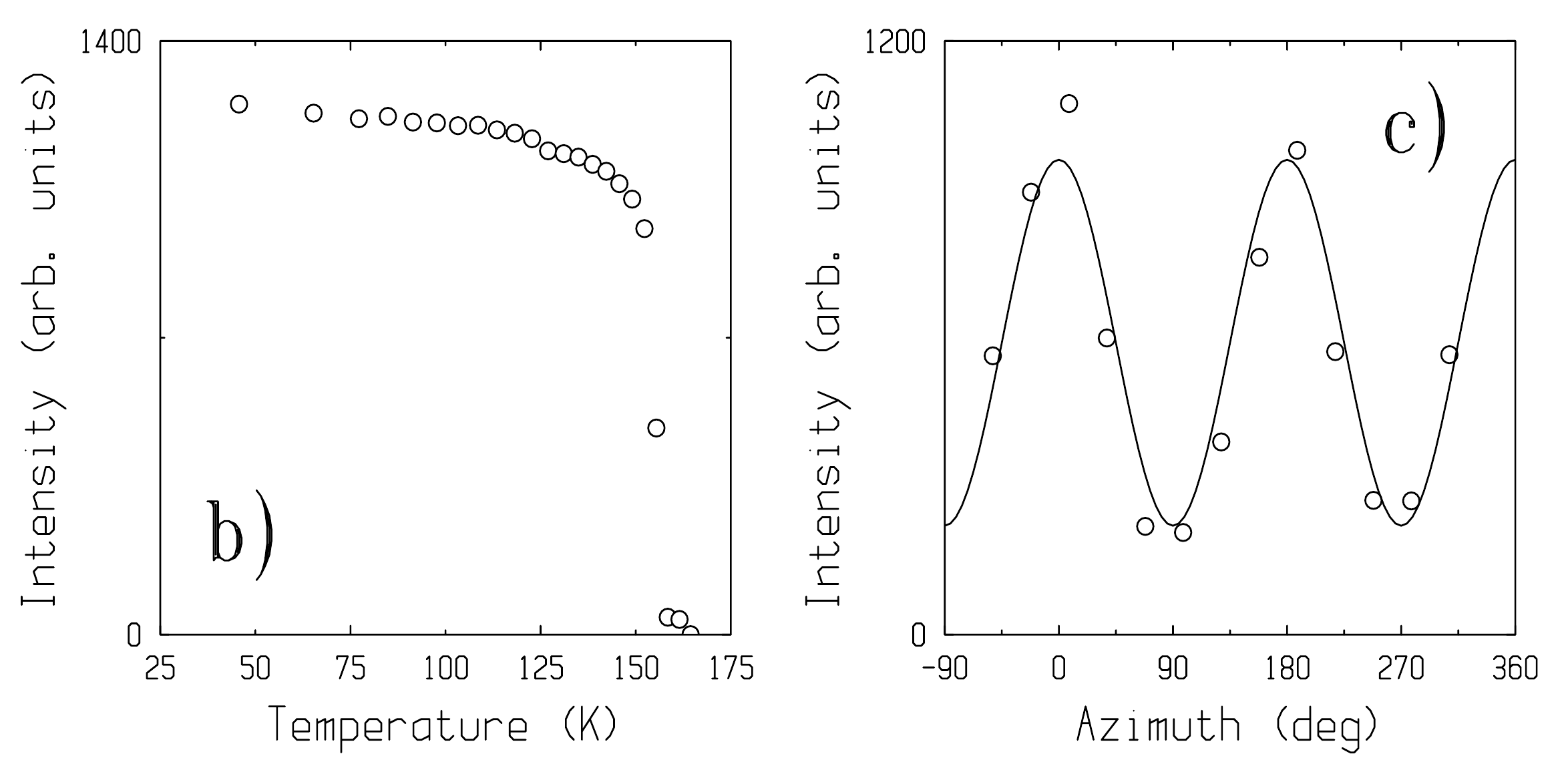}
\caption{\label{fig_LSMO} a) RXD spectra of the $(00\frac{1}{2})$
reflection at the O \emph{K}-edge of
La$_{\frac{7}{8}}$Sr$_{\frac{1}{8}}$MnO$_3$, with two incident
polarizations, $T=30$ K. $\psi\approx90$ b) Temperature dependence of
the reflection. c) Azimuthal dependence at 529.2 eV,
$\sigma\rightarrow\sigma+\pi$ channels, compared to a $\cos^2+\mathrm{const.}$
function (solid line), 0 deg. corresponds to the $\sigma$ incident
polarization along \textbf{b}, 90 deg. along \textbf{a}.}
\end{figure} 

\begin{figure}
\includegraphics[width=1\columnwidth]{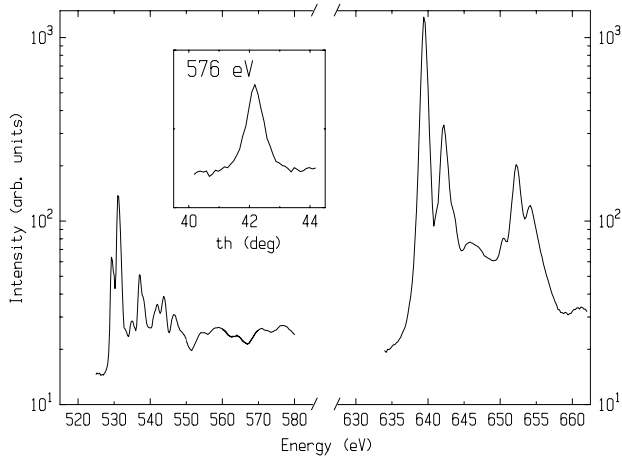}
\caption{\label{fig_LSMO2} Resonant diffraction spectrum of the
$(00\frac{1}{2})$ reflection at the O \emph{K}-edge and Mn $L$-edges of
La$_{\frac{7}{8}}$Sr$_{\frac{1}{8}}$MnO$_3$,
$\sigma\rightarrow\sigma+\pi$ channels, $T=30$ K. Three distinctive
regions should be considered. The amplitude at the O $K$-edge and Mn
$L$-edges is a probe of the states near $\varepsilon_f$, projected on O
and Mn atoms respectively. The amplitude between 535 and 560 eV is
mostly unsensitive to the active states, but reflects the positions of
neighboring atoms due to band structure effects.  Inset is the profile
of the reflection at 576 eV.}
\end{figure} 

We now turn to La$_\frac{7}{8}$Sr$_\frac{1}{8}$MnO$_3$.
Fig. \ref{fig_LSMO}a shows the RXD spectra of the $(00\frac{1}{2})$
superstructure reflection at the O $K$-edge for two incident
polarizations. Previous studies have already reported the unusual
stacking that doubles the unit cell along \textbf{c} below T$_{OO2}$
\cite{PhysRevLett.77.904}; the present data also reveals a specific
$2p$-state reordering (Fig. \ref{fig_LSMO}a and b). The fluorescence
yields are observed to be different for the two polarizations at the
edge, the differences providing evidence of an anisotropy of the
unoccupied O-states near $\varepsilon_f$. The main characteristic of the
$2p$ ordering is identified by considering the azimuthal dependence of
the reflection. The azimuthal scan never goes to zero; it has a near
perfect $\cos^2+\ \text{const}$ dependence (Fig. \ref{fig_LSMO}c). A
tensorial analysis of the resonant scattering permits to easily identify
the origin of the two terms \cite{Grenier04}: The constant term is due
to a net difference in the total charge carried by the oxygens on
different planes, whereas the sinusoidal dependence is due to the
relative direction of the polarization of the photon with the
anisotropic O electronic structure.

The diffraction vector $(0\ 0\ \frac{1}{2})$ corresponds to the net
difference in scattering, hence to a net difference in holes, from
planes separated by $\Delta z = 1$; a charge ordering at the $(0\ 0\
\frac{1}{2})$ vector provides a strong constraint to the theoretical
models. The models by Korotin \emph{et al.}  \cite{korotin.prb.00} and
by Geck \emph{et al.} \cite{geck:236401} both propose that hole-rich and
hole-poor Mn planes alternates along \textbf{c}, but with different
stacking orders. In Geck's model every other planes are identical when
projected on the \textbf{c} axis: they should not give the charge order
scattering we observe. In Korotin's model however, three successive
hole-rich planes alternate with one hole-poor plane. Such order should
give rise to a strong charge order signal on the O atoms because of the
hybridization, as we observe. We note that only Geck's model explains
reflections with a planar component observed elsewhere
\cite{geck:236401}, but the ordering along \textbf{c} is not consistent
with our data.

Finally, we also report on the scattering observed at this wave-vector
for energies well above the O $K$-edge (Fig. \ref{fig_LSMO2}). This
scattering probes the n$p$ continuum that is sensitive to the local
atomic configuration due to band structure effects. These features
signal presence of the lattice distortions accomodating the charge
ordering. Increasing again the photon energy, it was natural to look
also for the signal at the Mn $L$-edge which is again a direct probe of
the states nearby $\varepsilon_f$, but projected on the Mn atoms
(Fig. \ref{fig_LSMO2}). Correspondingly, the data are direct observation
of differing Mn $3d$ configurations; this will be discussed in detail
elsewhere.

In conclusion, resonant x-ray diffraction at the O $K$-edge has been
used to detect the electronic superstructures of the $2p$ states near
$\varepsilon_f$ in two manganites. In Bi$_{0.31}$Ca$_{0.69}$MnO$_3$, we
show evidence of an orbital ordering of hole-doped hybridized $2p-3d$,
and contradict O stripes with the same propagation vector as the Mn
orbital ordering. In La$_\frac{7}{8}$Sr$_\frac{1}{8}$MnO$_3$, the data
provided evidence of a charge ordering on the O atoms with the $(0\ 0\
\frac{1}{2})$ propagation vector, corresponding to a specific plane
stacking with differing hole concentrations. The orderings on O atoms
are not experimentally dissociated from the orderings on the Mn atoms,
in both compound, emphasizing the hybridized nature of the orbitals near
$\varepsilon_f$.

Quantification of the hole densities, and of the $2p-3d$ hybridization,
are now foreseeable. It requires \emph{ab initio} calculations of both
the ground states and the excited states, using dedicated formalisms, as
the excited state corresponds to an electronic structure with a strong
Coulomb potential in a core shell and an extra electron in the valence
shell. Finally, as the RXD technique can also be chemically-sensitive to
the magnetic moment, our results clearly suggest potential measurements
of magnetic superstructures on the oxygen atoms. The RXD data open a new
window on the $2p-3d$ states, the magnetic exchange couplings between
$3d$ metals and on the electronic superstructures at play in this
strongly correlated system.

We benefitted from the excellent support of the beamline staff on
X11MA-SLS, PSI. M.G.-F. thanks the Swiss National Science
Foundation. This work was also supported by US Dept of Energy, Division
of Materials Science, under contract No. DE-AC02-98CH10886. We thank
J. Debray of the Institut N\'eel for his excellent technical helps.


\bibliographystyle{apsrev}
\bibliography{../../../BiblioTex/manganites,../../../BiblioTex/cuprates,../../../BiblioTex/rxs}

\end{document}